\documentclass[showpacs,aps,amssymb,pre,floatfix]{revtex4}
\usepackage{graphicx} 
\usepackage{relsize}
\usepackage{mathtools}
\usepackage[overload]{empheq}

\begin{document}

\title{Phase ordering dynamics of reconstituting particles}

\author{F.\,A. G\'omez Albarrac\'in} 
\author{H.\,D. Rosales}
\author{M.\,D. Grynberg} 

\affiliation{IFLP-CONICET, Departamento de F\'isica, Facultad de Ciencias Exactas\\
Universidad Nacional de La Plata, C.C. 67, 1900 La Plata, Argentina}

\begin{abstract}
We consider the large-time dynamics of one-dimensional processes involving adsorption
and desorption of extended hard-core particles (dimers, trimers,\,$\cdots,k$-mers), 
while interacting through their constituent monomers. Desorption can occur whether 
or not these latter adsorbed together, which leads to reconstitution of $k$-mers and the 
appearance of sectors of motion with nonlocal conservation laws for $k \ge 3$. Dynamic 
exponents of the sector including the empty chain are evaluated by finite-size scaling 
analyses of the relaxation times embodied in the spectral gaps of evolution operators. 
For attractive interactions it is found that in the low-temperature limit such time scales 
converge to those of the Glauber dynamics, thus suggesting a diffusive universality class
for $k \ge 2$. This is also tested by simulated quenches down to $T=0$ where a common 
scaling function emerges. By contrast, under repulsive interactions the low-temperature 
dynamics is characterized by metastable states which decay subdiffusively to a highly 
degenerate and partially jammed phase.
\end{abstract}


\maketitle

\section{Introduction}

A variety of phenomena in physics and chemistry involving deposition of large particles,
such as colloids and proteins from a solution onto solid substrates, has long been 
investigated in terms of random sequential adsorption (RSA) models \cite{Flory,Evans}. 
In those processes, hard-core extended objects made of $k$-\,adjacent monomers 
arranged in a specified shape, drop onto a corresponding group of $k$-\,vacant sites 
randomly selected from a lattice substrate. Once a $k$-mer is accommodated it 
effectively blocks the available substrate area of all subsequent placements, so a 
limitting or jamming coverage rapidly emerges \cite{Evans}. However, recent 
experimental studies suggest that the actual roughness of the substrate may cause the 
detachment of a fraction of deposited colloids \cite{Shen}. In turn, other investigations 
evidence that detachment also plays a role in the kinetics of polymer chains at solid 
surfaces \cite{Frantz}, as well as in deposition of protein particles on DNA 'substrate' 
molecules \cite{DNA}. As a result, under small detachment rates the late kinetic stages 
of such processes are dominated by rearrangements of small empty areas into larger ones 
which, unlike ordinary RSA, can accommodate more particles and reach denser monolayer 
deposits \cite{Evans,Privman}.

Following the thread of ideas initiated in Refs.\,\cite{Barma1,Barma2}, here we further 
consider one-dimensional (1D) adsorption-desorption (AD) processes in which the 
detached $k$-mers do not necessarily correspond to the original deposited ones. In 
contrast to other relaxational models \cite{Evans} these processes contain no explicit 
monomer diffusion, but it is worth noting that the reconstruction of $k$-mers allows for 
an effective movement of these former, such as occurring e.\,g. in the AD sequence\, 
$\bullet \circ\circ\,\rightleftharpoons\,\bullet\bullet\bullet\,\rightleftharpoons\,\circ\circ 
\bullet$\,, say with dimers. Interestingly, for $k \ge 3$ these simple rules amount to a 
number of conservation laws that grows exponentially with the substrate size. At the root 
of this rather unusual partitioning of the phase space is a nonlocal construction, namely, 
the 'irreducible string' introduced by Barma and Dhar in Ref.\,\cite{Barma2}, and which 
we will briefly discuss in Sec.\,II. Thereupon, the question is whether the inclusion of 
interactions between adsorbates (fairly common in physisorbed species and self-assembly 
of nanoparticles \cite{Mazilu}), would affect the low-temperature dynamics at large-times  
when subject to these nonlocal conservations. 

Representing, as usual, monomers and vacancies respectively in terms of up and  down 
Ising variables, in what follows we shall think of this problem as a $k$-\,spin flip dynamics 
of an Ising chain with either ferro- or antiferromagnetic nearest-neighbor couplings. In the 
presence of detailed balance \cite{Kampen}, which we will assume throughout, such 
multi-spin processes may also be regarded as extensions of both Glauber ($k=1$) and 
Kawasaki dynamics ($k=2$) \cite{Glauber, Kawa}. Note that in the dimer case these 
AD processes can be readily reduced to a spin exchange kinetics, though with opposite 
coupling signs \cite{note1}. Thus, we shall focus on the situation $k \ge 3$ where the 
dynamics decomposes into many invariable sectors \cite{Barma2}. To ease the analysis 
we shall restrict ourselves to the so called 'null string' sector \cite{Barma2}, on the other 
hand the most common in the context of AD processes as it contains the initially empty 
substrate (cf. Sec.\,II).

As is known, in nearing the low-temperature limit the phase ordering dynamics of these 
1D processes become critical, being characterized by large relaxation times $(\tau)$ 
that grow with the equilibrium correlation length as $\xi^z$ \cite{Privman,Hohenberg}. 
Here, the dynamic exponent $z$ defines the universality class to which the dynamics 
belongs, and at late stages it basically describes how fast the length scale of the ordered 
phase is spreading after a quench from high temperatures \cite{Hohenberg,Puri}. In a 
finite system of typical length $L$, it is customary in practice to think of that scaling 
relation as a finite-size one, i.e. $\tau \propto L^z$, provided that $\xi$ becomes
comparable to the system size and this is taken sufficiently large \cite{Henkel}. Thus, in 
the following sections we shall exploit that finite-size approach to provide an estimation 
of dynamic exponents in the above reconstituting and interacting $k$-mer models. First, 
we will recast the master equation \cite{Kampen} governing those Markov processes 
in terms of a quantum spin representation of their associated Liouvillians or evolution 
matrices \cite{Kampen,Robin}. These latter lend themselves more readily to a finite-size 
scaling analysis of actual relaxation times as these are embodied in spectral gaps which 
we will subsequently evaluate by exact diagonalizations \cite{Lanczos,Dhar}. Also, for 
ferromagnetic (F) couplings we will complement those analyses with simulated quenches 
down to $T=0$.

However, when it comes to antiferromagnetic (AF) interactions this dynamics involves 
the passage through metastable states (see Sec.\,III), whose activation energy barriers 
($E_b$) make the nonequilibrium simulations difficult to implement at low-temperature 
regimes. Note that in that latter case, and regardless of the system being finite or 
not, the relaxation time scales can grow arbitrarily large owing to the contribution of 
Arrhenius factors $\propto e^{E_b/ T}$ \cite{note2}. Thereby when considering exact 
diagonalizations for the AF case, it will be appropriate to put forward a 'normalized' 
version of the above finite-size scaling hypothesis, namely
\vskip -0.35cm
\begin{equation}
\label{normal}
\lim_{T \to 0^+} e^{-E_b/ T}\,\tau \propto L^z\,,
\end{equation}
so as to ensure that $\tau$ is actually scaled within the 
Arrhenius regime. In fact, already in approaching that limit with our lowest accessible 
temperatures, a clear saturation trend of $e^{-E_b/T}\,\tau$ will be obtained for all sizes 
within reach. Also, and further to the case of AF interactions, it is worth anticipating here 
that for $k \ge 3$ a nontrivial and highly degenerate phase (rather than a plain AF state), 
will arise from the interplay between those couplings and the nonlocal conservations 
referred to above \cite{Barma2}.

The layout of this work is organized as follows. In Sec.\,II first we outline the basic 
transition probability rates of these processes along with their conservation laws which, 
irrespective of the presence of interactions and so long as all rates are kept nonzero, 
coincide with those of Ref.\,\cite{Barma2}. Then we exploit detailed balance to bring 
the evolution operator into a symmetric representation, thus simplifying the numerical 
analysis of Sec.\,III. In this latter, a sequence of finite-size estimates of dynamic 
exponents for $k = 3$ and 4 is obtained from the spectrum gaps of the associated 
quantum spin 'Hamiltonians'. Using standard recursive techniques \cite{Lanczos}, 
these are diagonalized within the subspaces of initially empty chains either with F or 
AF interactions. In addition, in the F case the scaling regimes of two-point correlations 
are examined for several values of $k$ after a quench from high temperatures. In the 
AF situation, statistical aspects of the exponentially degenerate ground state are also 
addressed. We close with Sec.\,IV which contains a summarizing discussion of results 
along with brief remarks on open issues and possible extensions of this work.

\section{Dynamics and conserved quantities}

The dynamics considered is set on a 1D lattice gas of $L$ sites each of which may be 
singly occupied (occupation numbers $n_i = 1$), or empty ($n_i = 0$). As usual in this 
context, the constituent particles (monomers) and vacancies are represented by the 
states of Ising spins $S_i = 2 n_i -1$ defining configurations $\vert S\, \rangle \equiv \vert 
S_1,\cdots,S_L \rangle$ of energies $E_S = - J\,\sum_i S_i\,S_{i+1}$. To account for either 
attractive or repulsive interactions between monomers, the coupling constant is set 
respectively as F or AF. To ease the subsequent discussion, henceforth we will assume 
that $L \propto k$ and that periodic boundary conditions (PBC) are imposed throughout.

Although Ising models have no intrinsic dynamics, a Markovian one can be prescribed 
with specific transition probability rates \cite{Glauber, Kawa}. These are thought of as 
stemming from energy fluctuations when the system is coupled to a heath bath at 
temperature $T$. For instantaneous quenches of this latter, the transition rates per unit 
time $W (S \to S' )$ between two configurations $\vert S\, \rangle, \vert S' \rangle$ (here 
differing in the state of $k$-\,consecutive parallel spins) are considered time independent 
and associated with the evolution operator ($H$) generating the dynamics via the matrix 
elements \cite{Kampen}
\begin{subequations}
\begin{empheq}[left={\langle\,S'\,\vert\,H \,\vert\,S\,\rangle = \empheqlbrace\,}]{align}
\label{ndiag}
& - W (S \to S')\,,\;\;\;\;\;\;{\rm for}\;\;S \ne S', \\
\label{diag}
& \sum_{S'\ne S}\, W (S \to S')\,,\;\;{\rm for}\;\;S = S'.
\end{empheq}
\end{subequations}
This evolution matrix allows to interpret the master equation \cite{Kampen} -governing 
the probability $P (S,t)$ to observe the system in a state $\vert P (t)\, \rangle = \sum_S 
P (S,t)\,\vert S\, \rangle$ at a given instant-  as a Schr\"odinger evolution in imaginary 
time, that is
\vskip -0.3cm
\begin{equation}
\label{master}
\frac{\partial}{\partial t}\, \vert P (t)\, \rangle = - H\, \vert P (t) \,\rangle\,.
\end{equation}
Hence by formal integration, the probability distribution at subsequent moments may 
be obtained from the action of $H$ on a given initial condition, i.e.  $\vert P (t) \,\rangle 
= e^{- H t}\,\vert P (0) \,\rangle$. In this regard, the first excitation mode of $H$ singles 
out the relaxation time of any observable as $\tau^{-1}={\rm Re}\;\lambda_1 > 0$,  
whereas by construction [\,Eqs.\,(\ref{ndiag}), (\ref{diag})\,], the steady distribution 
merely corresponds to a ground state with eigenvalue $\lambda_0 \equiv 0$. 

\begin{figure}[htbp]
\includegraphics[width=0.56\textwidth]{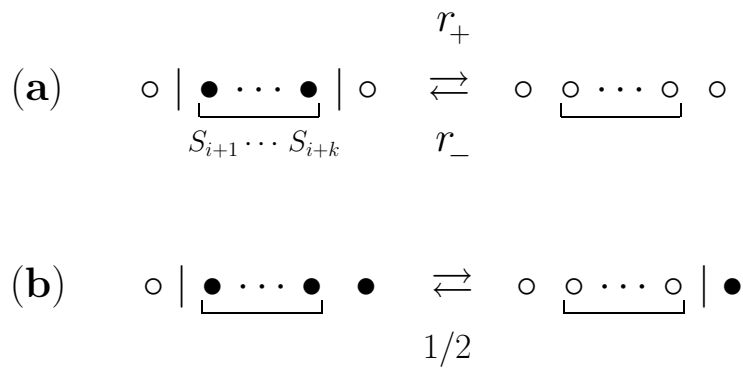}
\vskip -0.25cm
\caption{Schematic description of adsorbing-desorbing $k$-mers (within brackets), and 
their transition rates as defined in Eq.\,(\ref{rates}). Filled and empty circles denote spins 
with opposite orientations in turn forming kinks (vertical lines) on the dual chain. These 
former involve (a) pairing processes with rates $r_{\!_\pm }= \frac{1}{2} (\,1 \pm \tanh 
2K\,)$, as well as (b) hopping events with no energy changes.}
\label{processes}
\end{figure}

With respect to transition rates, these are set to satisfy the detailed balance condition 
\cite{Kampen}
\begin{equation}
\label{DB}
P_B (S) \,W(S \to S') = P_B (S') \, W(S' \to S)\,,
\end{equation}
so as to enforce the system to relax towards the Boltzmann distribution $P_B (S) \propto  
e^{-\beta E_S}$ at large-times. (In turn, detailed balance also allows for a symmetric
representation of the evolution operator; see Sec.\,II\,B). From now on temperatures are 
measured in energy units, or, equivalently, the Boltzmann constant in $\beta \equiv 
1/ (k_B T )$ is taken equal to unity. On the other hand, evidently there are many choices 
of $W$ that comply with Eq.\,(\ref{DB}). As usual in the context of kinetic Ising models 
\cite{Privman,Glauber,Kawa,Puri}, here we choose the Suzuki\,-\,Kubo form \cite{Suzuki}
\begin{equation}
\label{choice}
W (S \to S') = \frac{\alpha}{2}\Big\{\,1- \tanh \Big[\,\frac{\beta }{2} 
\,\big( E_{S'} - E_S \big) \Big]\,\Big\}\,,
\end{equation}  
with $1/\alpha$ just setting the time scale of the microscopic process and hereafter set 
to 1. In the specific case of the $k$-\,spin flip rates involved in the AD processes referred 
to in Sec.\,I, there are basically two situations in which the dynamics can proceed. These 
are depicted in Fig.\,\ref{processes} and later on will serve as a basis for the kink 
construction of Sec.\,II\,B.
\vskip -0.1cm

\noindent
As stressed before in Sec.\,I, all $k$-\,spin flips can take place whether or not their locations 
were previously flipped together, so the identity of desorbed $k$-mers is not necessarily 
preserved but is rather often reconstructed. Now, depending on the spin states neighboring 
the interval $[\,i\!+\!1\,, i\!+\!k\,]$ where these processes may eventually occur, and with 
the aid of the magnetization $m_{i,k} =  \frac{1}{k} \sum_{j=1}^k S_{i+j}$  associated 
to that region, the corresponding $W _{i,k}(S)$ rates derived from Eq.\,(\ref{choice}) can 
then be expressed as
\begin{eqnarray}
\label{rates}
W _{i,k}(S) &=& \frac{1}{2}\; \mathlarger{\delta}_{1,\, \vert m_{i,k} \vert}
\left[\,1- \,   \frac{m_{i,k}}{2}\, \big(\,S_i + S_{i+k+1} \big)\,
\tanh 2 K\, \right] \\
\nonumber
&=& \begin{cases}
\,r_{\!_\pm }= \frac{1}{2} (\,1 \pm \tanh 2K\,)\,,\;\;{\rm for}\;\;\Delta E = \mp 4 J\,,
\vspace{0.1cm} \cr
\,\frac{1}{2}\,,\;\;\;\;{\rm for}\;\;\Delta E = 0\,,
\end{cases}
\end{eqnarray}
where $K = \beta J$, and the Kronecker delta $\delta_{1,\, \vert m_{i,k} \vert} = 
\prod_{j=1}^{k-1}\,(1+ S_{i+j}\,S_{i+j+1}) /  2^{k-1}\!$ constrains all spins to be 
parallel within that interval. Unlike the Glauber dynamics where this latter difficulty 
does not arise, it is worth mentioning that the equations of motion set by such rates 
in Eq.\,(\ref{master}), generates a hierarchy which cannot be solved exactly. However, 
at least at the level of relaxation times (spectral gaps of $H$), let us anticipate that in 
the low-temperature limit all $\tau$'s become numerically indistinguishable  as long as 
$L/k$ is kept constant and F interactions are considered (see Sec.\,III\,A).

When it comes to conservation laws, since $L \propto k$  the lattice $\Lambda$ is 
$k$-\,partite in 1D (i.e. $\Lambda = \Lambda_1 + \cdots + \Lambda_k$), and so a set of 
independent constants of motion can be readily identified \cite{Barma1}. Since at every 
deposition (evaporation) step the number of incoming (outgoing) monomers is the same 
on each sublattice, then clearly their magnetization differences 
\begin{equation}
\label{m-diff}
M_a - M_b = \sum_{i \in \Lambda_a} S_i -  \sum_{i \in \Lambda_b} S_i\,,\;\;
a,b = 1,\cdots,k\,,
\end{equation}
will be maintained throughout the process. From these $k(k-1)/2$ differences, only $k-1$ 
of them are independent, so the number of conservation laws would grow at most as 
$L^{k-1}$. However, as mentioned in Sec.\,I, for $k > 2$ there is in fact a much more 
exhaustive set of constants of motion, in turn growing {\it exponentially} with the 
system size. 

\subsection{Irreducible strings}

\begin{figure}[htbp]
\vskip -0.05cm 
\includegraphics[width=0.65\textwidth]{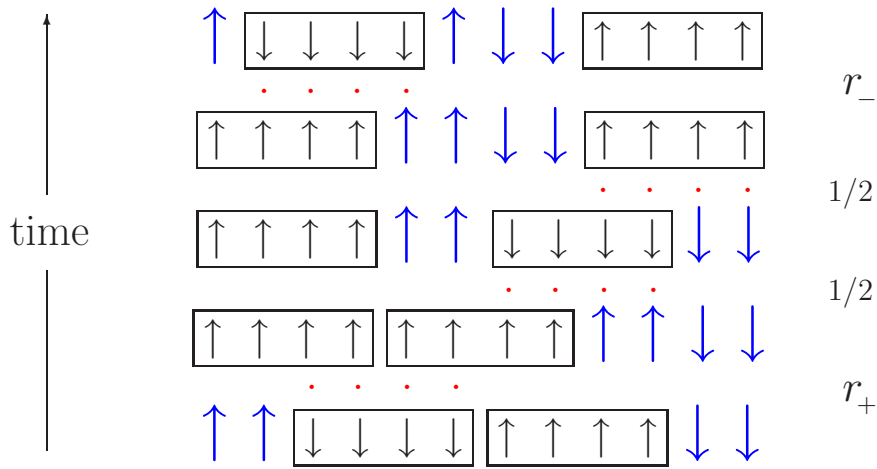}
\vskip -0.2cm
\caption{Schematic random walk of irreducible characters for $k=4$ and $L = 12$ with 
PBC's. The deletion procedure described in the text is here depicted by the boxes around 
reducible groups of spins. At each step the identification of these latter is made, say, 
from left to right. Larger spins denote the irreducible characters whose ordering is left 
invariant by the dynamics. Dots signal the locations of updated spins with rates indicated 
at the rightmost column (see Eq.\,(\ref{rates}) and Fig.\,\ref{processes}).}
\label{walk}
\end{figure}

To understand that latter issue, following Ref.\,\cite{Barma2} we now define the 
irreducible string (IS) $I \{S_1,\cdots,S_L\}$ of a given spin configuration as the 
sequence obtained by deleting all groups of $k$ consecutive parallel spins appearing
on chosen locations, and then repeating recursively the procedure on the resulting 
shorter string ($\propto k$) until no further such groups remain. As an illustration, 
consider the following examples, say for $k=4$,
\begin{eqnarray}
\nonumber
I \big\{\!\uparrow \downarrow \fbox{$\!\uparrow \uparrow \uparrow \uparrow\!$}
\uparrow \uparrow\! \big\} &=& I \big\{\!\uparrow \downarrow \uparrow\uparrow
\fbox{$\! \uparrow\uparrow \uparrow\uparrow\!$}\,\big\} = \big\{\!\uparrow 
\downarrow \uparrow \uparrow \!\big\},\\
\label{null}
I\,\Big\{\,\fbox{$\uparrow \uparrow\!$ \fbox{$\!\downarrow \downarrow \downarrow 
\downarrow\!$}\,$\uparrow \uparrow$}\,\Big\} &=& \big\{\,\emptyset\,\big\},
\;({\rm null\,\, string}),\\
\nonumber
I \big\{\!\uparrow \uparrow \downarrow \downarrow \uparrow  \downarrow \downarrow 
\downarrow \!\big\} &=&   \big\{\!\uparrow \uparrow \downarrow \downarrow \uparrow 
\downarrow \downarrow \downarrow\!\big\},\;({\rm full\;\, jammed})\,.
\end{eqnarray}
In the first case, this deletion (marked by boxes) is applied to a group of spins chosen 
starting from either the left or right, the actual location of the targeted group being 
irrelevant. In the second instance the procedure is carried out recursively in two steps 
and no characters are left. In the third example the string considered is already jammed 
and cannot evolve further. The invariance of the irreducible characters (if any) left by this 
process is in line with the idea that successive AD attempts on a given spin configuration 
just changes the position of those characters by multiples of $k$ lattice spacings. The 
separations between them are mediated by substrings of different lengths $\propto k$,  
though all of these are in turn reducible to null strings ({\bf NS}) \cite{Barma2}. Thus, 
the AD dynamics may be thought of as a random walk of hard-core irreducible characters 
(they cannot cross each other), as depicted schematically in Fig.\,\ref{walk}. The positions 
of these walkers at a given instant of course depend on the order in which the reduction rule 
is applied, but the issue to bear in mind here is that the order in the sequence of irreducible 
characters remains unaltered throughout.

Due to the highly convoluted form in which that sequence is obtained, it is clear that 
$k$ parallel spins well separated from each other may or may not form a reducible 
$k$-mer depending on the substrings in between. In that sense, the IS conservation 
is non-local as it involves the whole configuration. More importantly, as mentioned in 
Ref.\,\cite{Barma2} note that two spin configurations $\vert S \rangle,\,\vert S' \rangle$ 
are connected by the dynamics $\Longleftrightarrow I \{S\}= I \{S '\}$. As long as all AD 
rates are held finite, evidently this also applies in the presence of interactions at $T > 0$. 
Thus, the IS uniquely labels all subspaces left invariant by the $k$-mer dynamics, and
regardless in which order the reducible groups are removed. In particular, for $k > 2$ 
the AF state is therefore {\it unreachable} either from initially empty or closed packed 
chains as these latter belong to NS sectors, whereas an AF configuration is already 
irreducible. We will further discuss this issue in Sec.\,III\,C when characterizing the actual 
ground state in NS spaces for $J < 0$.

Now it is clear that the number of combinations forming all irreducible sequences for 
$k >2$ grows exponentially with the number of characters or string length ${\cal L}\le 
L$ (with ${\cal L} \propto k$). More specifically, a straightforward analysis of a recursive
relation for this $\cal L$-\,dependent quantity \cite{Barma1,Barma2}, shows that for large 
${\cal L}$ and $k > 2$ the number of invariant subspaces or string sectors increases as 
fast as $x^{\cal L}\!$, with $x$ being the largest root of $x^k= 2\,x^{k -1} - 1$. As for 
dimers, where this AD process reduces to a spin exchange kinetics \cite{note1}, it is 
worth pointing out that although the IS construction is still well defined, in this case all 
strings simply consist of AF sequences of lengths ${\cal L} = L, L-2, \cdots, 2,0$; each 
one (except ${\cal L} = 0$) having two possible orientations. So, these $L+1$ strings 
just correspond to each of the $\pm \cal L$ total magnetizations left invariant by the 
Kawasaki dynamics.

\subsection{Kink representation}

Turning to the evolution operator, it is convenient at this point to move to the domain 
wall or kink representation, so as to halve the dimensions of the stochastic matrix in 
the diagonalizations of Sec.\,III. In that two-to-one mapping (outlined before in 
Fig.\,\ref{processes}), new Ising variables $\sigma_i \equiv - S_i \, S_{i+1}$ with energy 
$E_{\sigma} = J\sum_i \sigma_i$\, stand on dual chain locations where pairing and diffusion 
of kinks ($\sigma = 1$) may take place. Such dual processes can then be schematized as
\vspace{-0.15cm}
\begin{subequations}
\begin{eqnarray}
\nonumber
& r_{\!_+ } & \\
1 \; \underbracket[0.4pt]{\circ \cdots\,}_{k-1}\; 1\;\;  
&  \underbracket[0pt]{\rightleftarrows}_{\mathlarger r_{\!_- }} & \;\;  \circ\;\, 
\underbracket[0.4pt]{\circ \cdots\,}\;\circ\;\;\;\;\,
({\rm pairing}\,,\;\Delta E = \mp 4 J)\,,\\
\nonumber\\
1 \; \underbracket[0.4pt]{\circ \cdots\,}_{k-1}\;\circ\;\,
& \underbracket[0pt]{\rightleftarrows}_{1/2} & \;\; \circ \;\, 
\underbracket[0.4pt]{\circ \cdots\,}\, 1\;\;\;\;\;
({\rm diffusion}\,,\;\Delta E = 0)\,,
\end{eqnarray}
\end{subequations}
with all brackets involving $k\!-\!1$ vacancies ($\sigma =\!-1$) in the dual chain. 
Now, thinking of these kink configurations as 1/2-\,spinors states $\vert\,\sigma\,\rangle 
\equiv \vert\,\sigma_1,\cdots \sigma_L \,\rangle$ (say in the $z$ direction), clearly these 
processes can be associated with the action of usual raising and lowering operators 
$\sigma^+,\sigma^-$. More specifically, introducing the projectors
\begin{equation}
\label{proj}
\hat P_i^{(k)} = \frac{1}{2^{k-1}}\;\prod_{j=1}^{k-1}\, (1 - \sigma^z_{i+j})\,,
\end{equation}
[\,here playing the role of the Kronecker delta in Eq.\,(\ref{rates})\,], so as to allow
transitions mediated only by empty intervals $[\,i\!+\!1\,, i\!+\!k-1\,]$, then the 
operational nondiagonal counterparts of Eq.\,(\ref{ndiag}) accounting for both 
pairing and diffusion events sketched in (9), are respectively given by
\begin{subequations}
\begin{eqnarray}
\label{pair}
H_{nd}^{^{(pair)}} &=& - \sum_i\,\hat P_i^{(k)} \left(\, r_{\!_- }\,\sigma^+_i\, 
\sigma^+_{i+k}\,+\, r_{\!_+ }\,\sigma^-_i\,\sigma^-_{i+k}\,\right)\,,\\
\label{diff}
H_{nd}^{^{(di\!f\!f)}} &=& -\frac{1}{2}\,\sum_i\, \hat P_i^{(k)} \left(\,
\sigma^+_i\,\sigma^-_{i+k} + {\rm H.c.}\, \right)\,.
\end{eqnarray}
\end{subequations}

Although this would leave us with a nonsymmetric evolution operator, we can now exploit 
detailed balance to bring Eq.\,(\ref{pair}) into a symmetric form. In this context this 
amounts to considering the nonunitary spin rotation 
\begin{equation}
\label{rot}
R = \exp \Big(-i \frac{\varphi}{2} \sum_j \sigma^z_j\,\Big)\,,
\end{equation}
\vskip -0.2cm
\noindent
with pure imaginary angles $\varphi = i K$. Since $R$ is diagonal and its matrix elements 
$\langle\, \sigma\, \vert\, R\,\vert\, \sigma \rangle = e^{\,\frac{\beta}{2} E_{\sigma}}$ 
just involve the above kink energies, then all transition rates derived from 
Eq.\,(\ref{ndiag}) will transform as
\begin{equation}
W ( \sigma \to \sigma') \to e^{\,{\frac{\beta}{2}}\, 
\left(E_{\sigma'} - E_{\sigma} \right)}\, W ( \sigma \to \sigma')\,.
\end{equation}
Taking into account that $W ( \sigma \to \sigma')$ also comply with detailed balance 
[\,Eq.\,(\ref{DB})\,], it is thus clear that in the transformed representation all these 
nondiagonal elements become symmetrical, i.e. $ W ( \sigma \to \sigma') \to [\,W 
(\sigma \to \sigma') \;W ( \sigma' \to \sigma)\,]^{1/2}$. In particular, under the spin 
rotation (\ref{rot}) the pairing terms of Eq.\,(\ref{pair}) transform as
\begin{equation}
r_{\!_\mp }\, \sigma^{\pm}_i\,\sigma^{\pm}_{i+k} \, \to \,
 \frac{1}{2}\;{\rm sech}\,2K \; \sigma^{\pm}_i\,\sigma^{\pm}_{i+k}\,,
\end{equation}
while leaving projectors (\ref{proj}) and all diffusion terms of Eq.\,(\ref{diff}) unchanged.
Thus, collecting the contributions of $H_{nd}^{^{(pair)}}\!\!+H_{nd}^{^{(di\!f\!f)}}\!
\!\!$, the symmetrized operational analog of Eq.\,(\ref{ndiag}) is now given by
\begin{eqnarray}
\label{sim}
H_{nd}^{(k)} = -\frac{1}{4}\,(1+ {\rm sech}\,2K)\,\sum_i\,\hat P_i^{(k)} \left(\,  
\sigma^x_i\, \sigma^x_{i+k} + \tanh^2\!K\;\sigma^y_i\, \sigma^y_{i+k} \,\right)\,.
\end{eqnarray}

To complete the construction of the evolution operator, we last turn to the diagonal 
matrix elements of Eq.\,(\ref{diag}) needed for conservation of probability. By definition, 
these elements count the number of weighted manners in which a given configuration 
$\vert\,\sigma\,\rangle$ can evolve to different ones either in one pairing or diffusion step. 
As before, this can be tracked down in terms of the above $\hat P_i^{(k)}\!$ projectors, 
while concurrently probing the appropriate occupancy of kinks and vacancies at sites 
$i, i+k$. So, in adding those diagonal contributions the counterpart of Eq.\,(\ref{diag}) 
becomes
\begin{equation}
H_d^{(k)} = \frac{1}{2}\,\sum_i\,\hat P_i^{(k)} \Big[\, 1 + \,\frac{1}{2}\,
\tanh 2K\, \big( \sigma^z_i + \sigma^z_{i+k}  \big)\,\Big]\,,
\end{equation}
which in turn is left invariant by the spin rotation (\ref{rot}). It is worth noting that 
for $k=1$ (Glauber dynamics) no projectors are necessary and $H_{nd} + H_d$ fully 
recovers the bilinear form of Ref.\,\cite{Glauber}, being ultimately reducible to a free 
fermion Hamiltonian. By contrast, for $k>1$ projectors (\ref{proj}) introduce correlated 
pairing and hopping terms as well as many-body interactions, in which case the evolution 
operator is no longer soluble (cf. however Fig.\,\ref{four}a in Sec.\,III\,A).

As for the string sectors obtained from the reduction rules \cite{Barma2} summarized in 
Sec.\,II\,A, it can be readily checked that in the dual representation those reductions amount 
to deleting $k$-\,contiguous vacancies (\,$\circ \cdots^{\!\!\!\!\!^k}\;\circ \to \emptyset$\,), 
along with replacing kink pairs with $k\!-\!1$ vacancies in between by just one vacancy (\,$1 
\circ \cdots^{\!\!\!\!\!\!\!\!\!^{k-1}} 1 \to \circ$\,). As a result of the repeated applications of 
these reductions one is left with string sequences of length ${\cal L} \propto k$ where there 
can be no more than $k-2$ consecutive vacancies, each sequence being left invariant by 
the dual dynamics. In practice, to deploy a complete set of NS configurations (where the
diagonalizations of Sec.\,III are carried out), we shall successively apply Eq.\,(\ref{sim}) to 
the states stemming from an initial one in that sector until this latter is exhausted.

\section{Numerical results}

Armed with $H^{(k)} \equiv H_{nd}^{(k)} + H_d^{(k)}$ acting on generic kink states, 
we can now implement a Lanczos diagonalization procedure \cite{Lanczos} without having 
to store in memory the matrix representation of the evolution operator. As mentioned in 
Sec.\,I, we focus on the NS sector ($S_{\emptyset}$) containing the initially empty chain 
and restrict ourselves to the cases of $k=3$ and 4. Firstly, as a consistency check, we 
verified that transforming the Boltzmann distribution $R\,\vert P_B \rangle \propto 
\sum_{\sigma} \exp( -\frac{\beta}{2} E_{\sigma}) \vert\,\sigma \rangle$ with rotation 
(\ref{rot}) and $\vert \sigma\rangle \in S_{\emptyset}$, actually produces a ground state 
of $H^{(k)}$ with eigenvalue $\lambda_0 \equiv 0$. This also served to start up the 
Lanczos recursion but with a random initial state  chosen orthonormal to that equilibrium 
direction. In turn, all subsequent states generated by the Lanczos algorithm were also 
reorthogonalized to $R\,\vert P_B \rangle$. Thereafter, we obtained the first excited 
eigenmodes of $H^{(k)}$ for lengths of up to $L=30$ for $k=3$ and $L =36$ for $k=4$, 
the main limitation for this being the exponential growth of the space dimensionality in 
$S_{\emptyset}$ \cite{Barma2}.

\subsection{Ferromagnetic case}

In this situation, after a quench to low-temperature regimes the large-time dynamics of 
$S_{\emptyset}$ is essentially mediated by kinks that diffuse at no energy cost (rate 
1/2). Owing to the projectors of Eqs.\,(\ref{pair}) and (\ref{diff}) however note that kinks 
can not cross each other, neither annihilate in the presence of other kinks in between, nor 
if they diffuse through different sublattices. But taking into account the reduction rules 
referred to in Sec.\,II\,B for the dual representation, there must be at least one interval 
with two kinks separated by $nk - 1$ vacancies ($1 \le n \le \frac{L}{k} - 1$); otherwise 
the explored configurations would not be fully reducible to $S_{\emptyset}$. Hence, 
there are regions where kinks can always meet at a distance of $k$ lattice spacings and 
annihilate in pairs ($\Delta E = -4 J$) with rate $r_{\!_+} \lesssim 1$, which in the limit 
of $T \to 0$ gives rise to a monotonic coarsening of F domains. Therefore, close to the 
equilibrium regime of $S_{\emptyset}$ the characteristic time involved between pair 
annihilations is that for kinks to diffuse across a correlation length $\xi \propto L$, so a 
relaxation time $\tau$ growing as $L^2$ might be expected. 

\begin{figure}[htbp]
\hskip -7.5cm
\includegraphics[width=0.43\textwidth]{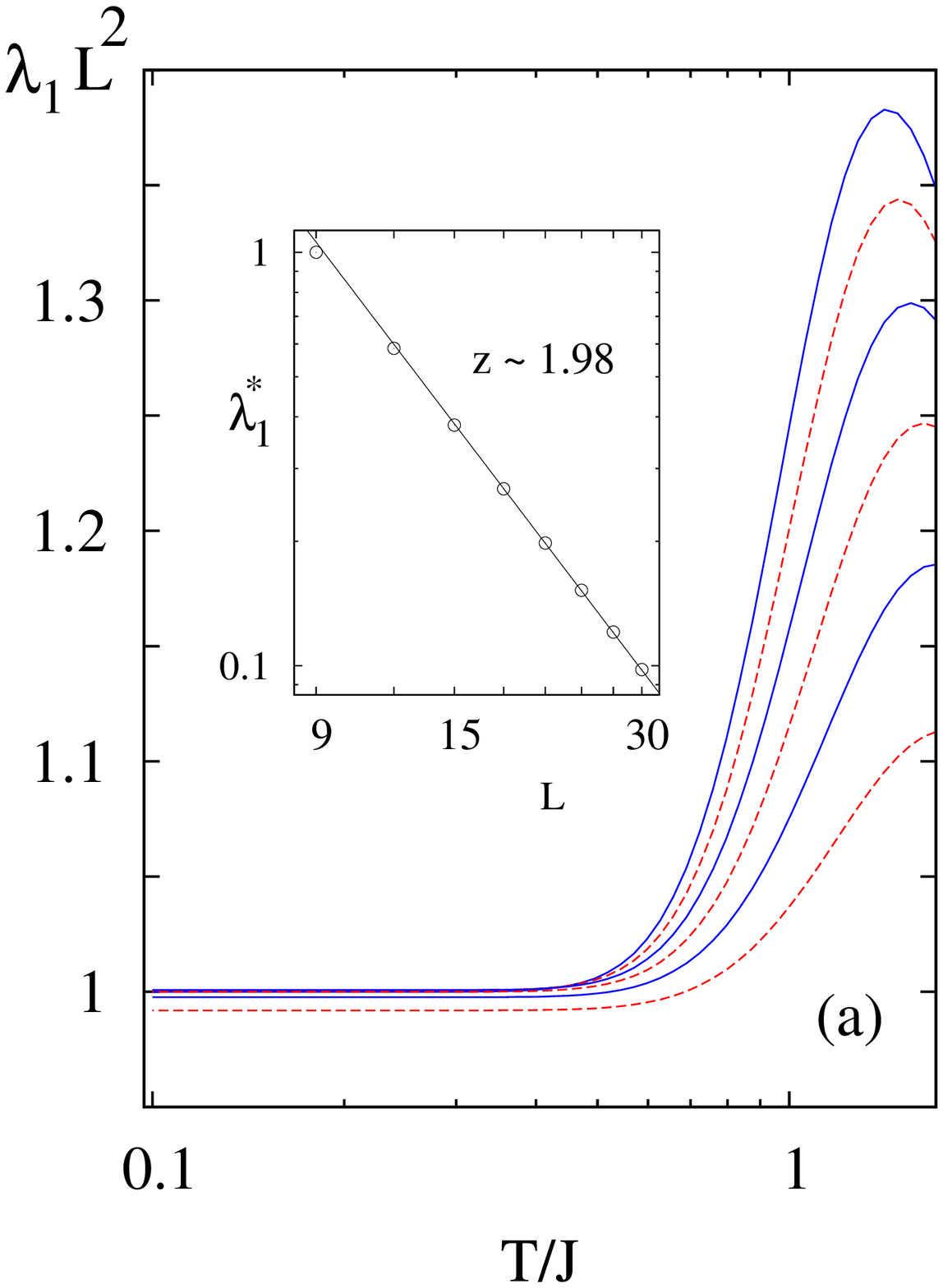}
\vskip -8.73cm
\hskip 7.5cm
\includegraphics[width=0.43\textwidth]{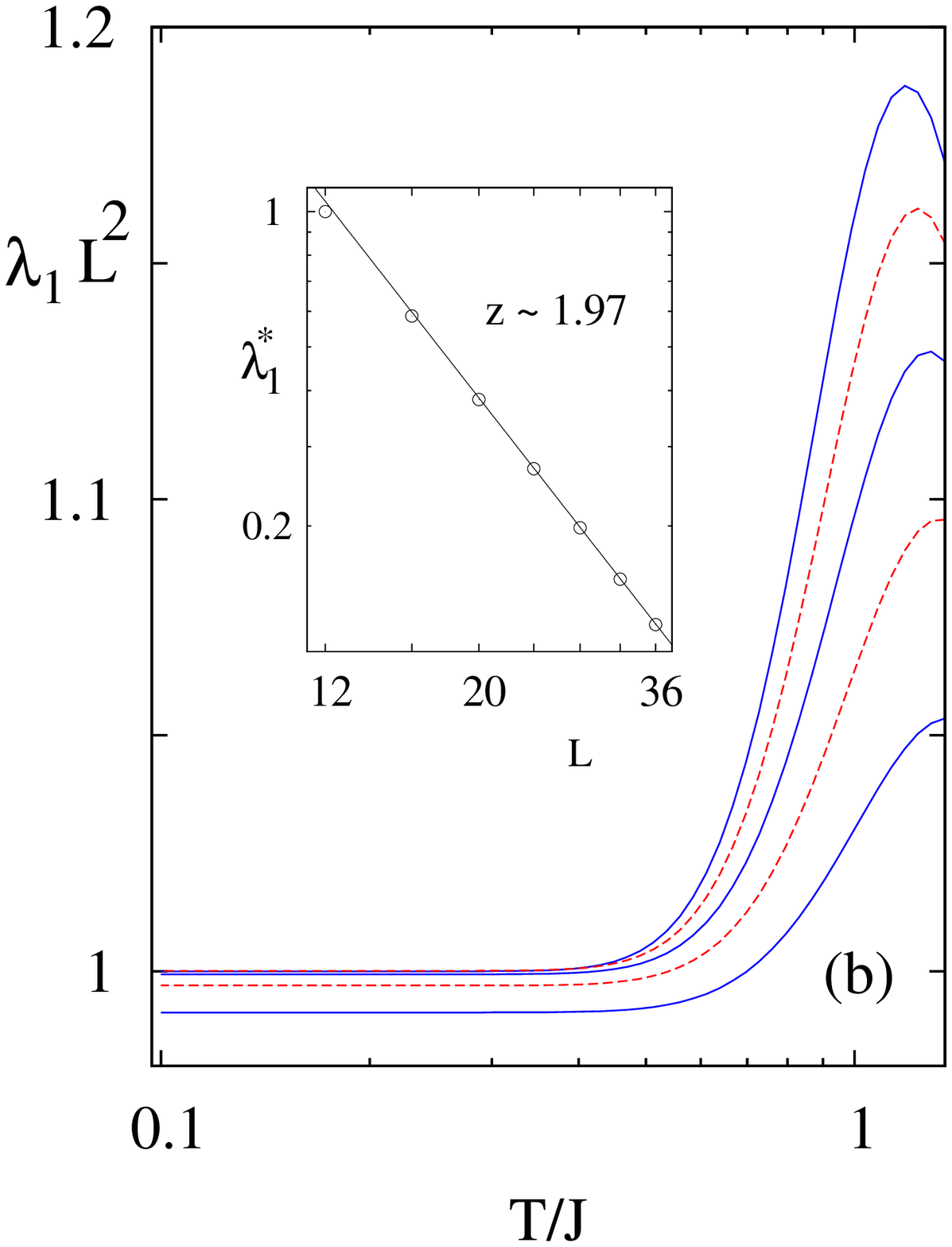}
\vskip 0.25cm
\caption{Finite-size scaling of spectral gaps of the evolution operator ($J > 0$) on 
approaching low-temperature regimes in NS sectors for (a) $k=3$, and (b) $k=4$. From 
top to bottom solid and dashed lines stand in turn for sizes $L \propto k$ with (a) $L = 30,
\cdots,15$, and (b) $L = 36,\cdots,20$. The slopes of the insets estimate the typical 
finite-size decay of these gaps in the limit of $T \to 0$, thus suggesting a common 
dynamic diffusive exponent. For displaying convenience, in (a) and (b) the vertical scales 
of main panels were normalized by factors $30^2\, \lambda_1^{\!^*}(30)$ and $36^2\, 
\lambda_1^{\!^*} (36)$ respectively.}
\label{FSS}
\end{figure}

\begin{figure}[htbp]
\vskip -0.2cm
\hskip -8cm
\includegraphics[width=0.44\textwidth]{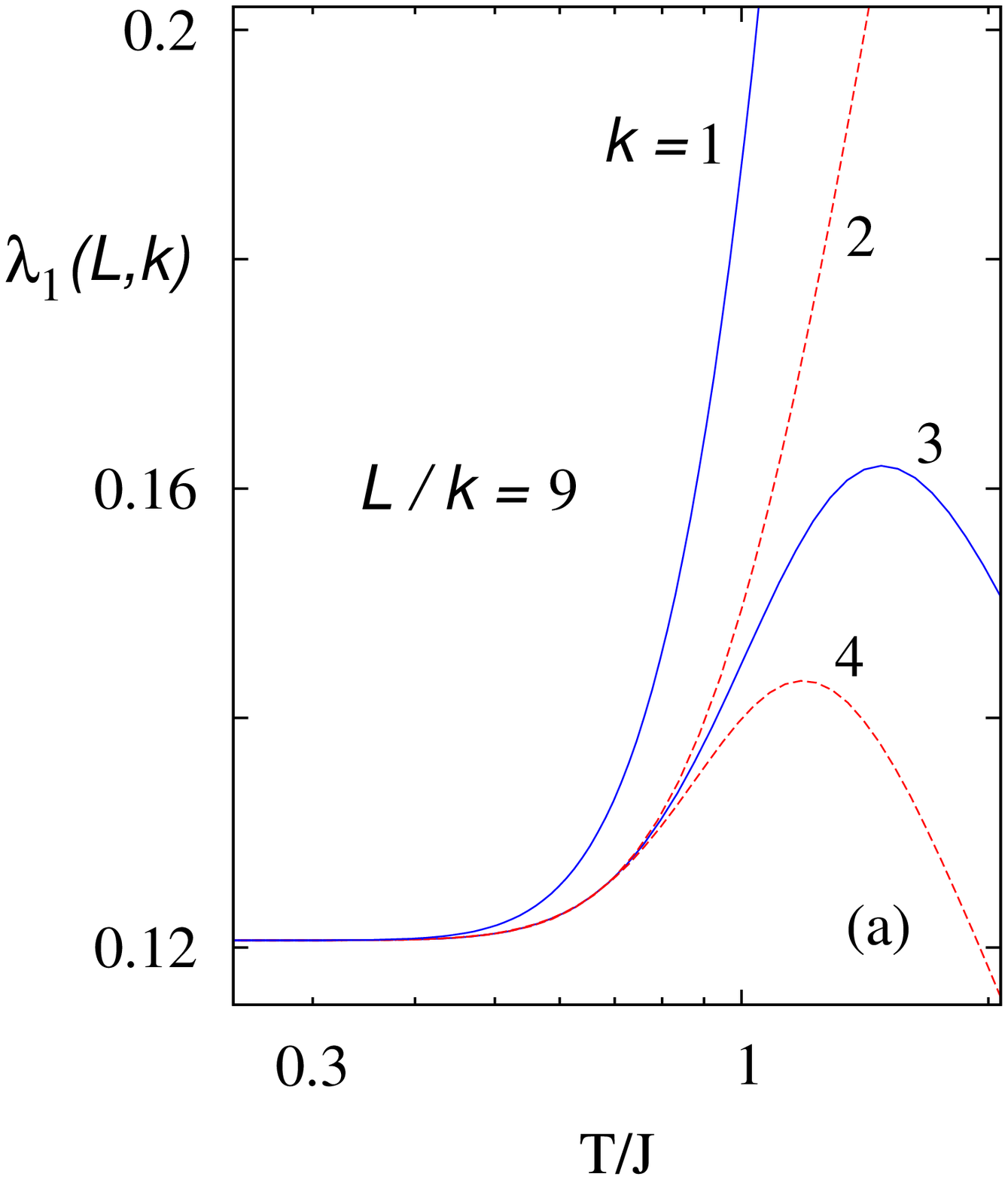}
\vskip -8.6cm
\hskip 8cm
\includegraphics[width=0.414\textwidth]{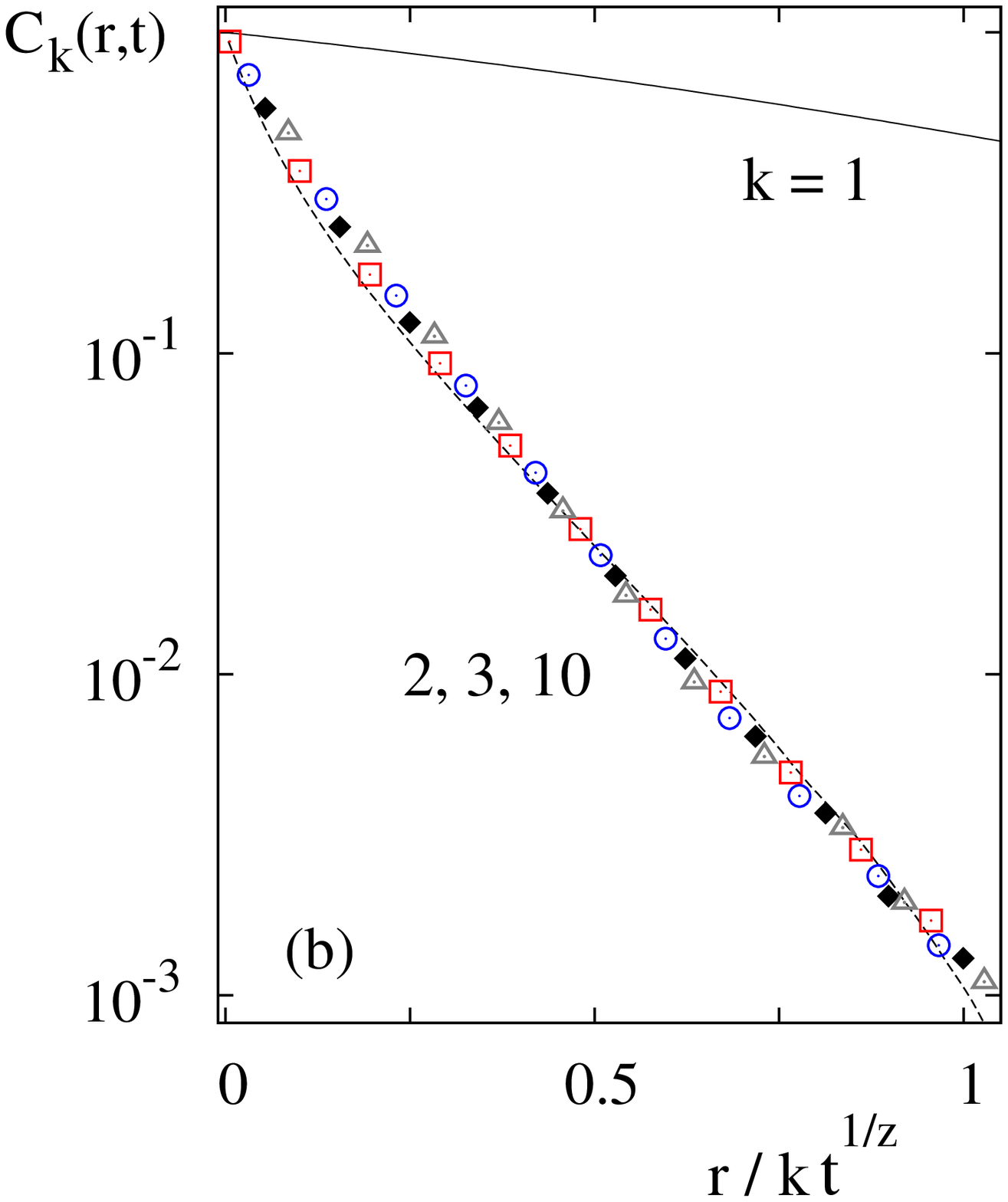}
\vskip 0.2cm
\caption{(a) Low-temperature scaling of spectral gaps ($J > 0$) for several $k$-mer 
sizes. As $T \to 0$, all gaps collapse onto the exact solution of the monomer case 
(Glauber dynamics), provided the chain length of this latter is rescaled by a factor 
$1/k$. Other reachable values of $L/k$ (not shown) also follow the Glauber solution in 
that limit. (b) Scaling plot of spin-spin correlation functions ($J > 0$) for $L = 1.2 \times 
10^5$ after a quench down to $T=0$. The data collapse was attained upon renormalizing 
$r$- distances as $r/k$ while choosing dynamic exponents $z \in (1.96\,,\,2.07)$. Squares 
($t =5 \times 10^3$ steps) and circles ($t = 5 \times\!10^4$) depict the case $k=3$,
whereas rhomboids ($t = 10^3$) and triangles ($t = 10^4$) stand for $k=10$. Error bars
are smaller than twice the symbol sizes. The dashed line represents the case $k=2$ after 
$10^4$ steps, while the solid one denotes the exact scaling function of Ref.\,\cite{Bray}.}
\label{four}
\end{figure}

In fact, this is evidenced in Figs.\,\ref{FSS}(a) and \ref{FSS}(b) where scaling plots 
of the spectral gaps ($\lambda_1 = 1/\tau$) of $H^{(k)}$ are shown for $k=3$ and 4 
respectively. As temperature is lowered the data collapse towards larger sizes is attained 
on choosing a diffusive dynamic exponent ($z=2$). This is also in close agreement with 
the slopes read off from the insets, in turn estimating these finite-size gaps in their low 
temperature limit [\,$\lambda_1^{\!^*}(L) := \lim_{\,T \to 0}\, \lambda_1 (L)$\,].
Moreover, let us now consider Fig.\,\ref{four}(a) and compare the spectral gaps of the 
Glauber dynamics \cite{Glauber}, i.e. $\lambda_1 (L) = 2\, \big( 1 - \tanh 2 K\, \cos \frac
{\pi}{L} \big)$, with those resulting from our finite-size diagonalizations for $k=2, 3$, and 
4. Interestingly, it turns out that in lowering the temperature $\lambda_1 (L, k)$ converges 
towards the exact solution of $k=1$ so long as the length of this latter case is rescaled as 
$L/k$, i.e.  $\lambda_1^{\!^*} (L, k)= 2\, \big( 1 - \cos\frac{ \pi k}{L} \big)$. In particular, 
for $T/J \alt 0.25$ we checked out that for all sizes in reach these gaps become numerically 
indistinguishable, at least within quadruple precision. Therefore, as far as the critical 
dynamics is concerned, when $L \gg k$ these comparisons strongly suggest that in the 
NS sector the fundamental scaling relation between $\tau$ and $L$ is just a diffusive one; 
more specifically
\begin{equation}
\label{exact}
\tau = \Big(\frac{L}{\pi k}\Big)^2 + {\cal O} (1)\,.
\end{equation}
In passing note that for $k=2$ where, as mentioned before, the problem reduces to a 
Kawasaki dynamics with $J < 0$ (no metastability), this diffusive picture is consistent 
with the results encountered in previous studies \cite{Lage}.

Now we check whether this length resizing is of any consequence on a more microscopic 
level of description, such as the equal-time two-point correlators $C_k (r,t) = \frac{1}{L} 
\sum_j \langle S_j\, S_{j+r} \rangle (t)$ in the original spin representation. In Fig.\,\ref{four}
(b) we display these functions in the NS sectors of several $k$-mers by simulating quenches 
down to $T=0$ from initially disordered states. 
For $k > 2$ the sampling of these latter poses the nontrivial problem of generating an 
equally weighted distribution of NS configurations. As an approximation though, here each 
disordered sample was obtained by evolving the AD process through $10^6$ steps in the 
high temperature limit ($J = 0$), starting from $L/k$ consecutive $k$-mers randomly 
oriented (particle or vacancy).  On par with Fig.\,\ref{four}(a), it turns out that after 
averaging over $\sim 3 \times 10^3$ independent samples in chains with $1.2 \times 
10^5$ sites, there is a common scaling form into which these correlators can be made to 
collapse provided all spin separations are rescaled as $r/k$,\, i.e. $C_k (r,t) \simeq F \big(
r /\; k \,t^{1/2} \big)$. For monomers, where there is no equivalent to the IS construction 
neither conserved quantities, an exact scaling form for these correlators can be cast in 
terms of the complementary error function, namely \cite{Bray}
\vskip -0.3cm
\begin{equation}
C_1(r,t) = {\rm erfc}\big(\, r/\;2\, t^{1/2}\big)\,,
\end{equation}
although as is shown in Fig.\,\ref{four}(b), it is rather apart from the other $k$-mer 
cases. Nonetheless, and in line with Eq.\,(\ref{exact}), clearly all these AD situations 
are characterized by ferromagnetic length scales coarsening as $k\,t^{1/2}$.

\phantom{}
\vspace{-1.5cm}

\subsection{Antiferromagnetic case}

\begin{figure}[htbp]
\vskip 0.2cm
\includegraphics[width=0.45\textwidth]{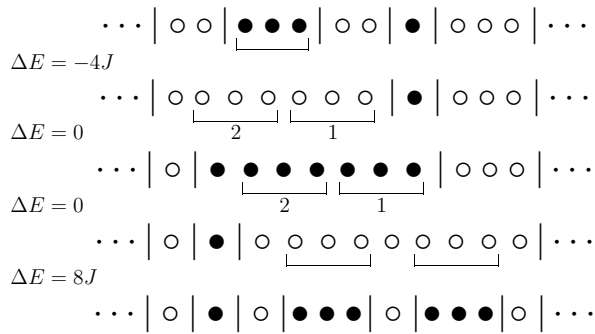}
\vskip 0.25cm
\caption{Schematic decay from a trimer metastable state (top) under $J < 0$. After 
surmounting the initial energy barrier, kinks separating opposite spin orientations can 
diffuse freely until an interval of nine parallel spins shows up. This allows to decrease 
the initial energy by creating two kink pairs (bottom). Brackets show updated locations, 
whereas numbers in diffusion events indicate the ordering of those updates.}
\label{decay}
\end{figure}

By contrast to the F dynamics, under AF interactions often this system can reach states in 
which further energy-lowering processes are unlikely at low temperature regimes. These 
configurations are such that there can be no more than $k$-\,consecutive parallel spins, 
and proliferate exponentially with $L$. Although the AF dynamics attempts to maximize 
the number of kinks, note that once one of these states is reached there is no way to escape 
from it without first annihilating a kink pair ($\Delta E = - 4 J$). If that annihilation in turn 
produces at least $2 k+1$ parallel spins, then by subsequent diffusion of kinks eventually a 
domain of $2k +3$ spins can be accommodated at no further energy cost. A scheme of this 
decay process, say for $k=3$, is depicted in Fig.\,\ref{decay}. 
\noindent 
This prepares the conditions to create {\it two} kink pairs with which the original 
metastable energy is finally lowered in $\Delta E=4 J$. As indicated in Fig.\,\ref{decay} note 
that this also requires the initial presence of at least two $k$-mer locations, no matter how 
distant they might be \cite{note3}. Later on we will make use of this decay pattern when 
sampling ground states of larger chains (Sec.\,III\,C).

\begin{figure}[htbp]
\hskip -7.5cm
\includegraphics[width=0.43\textwidth]{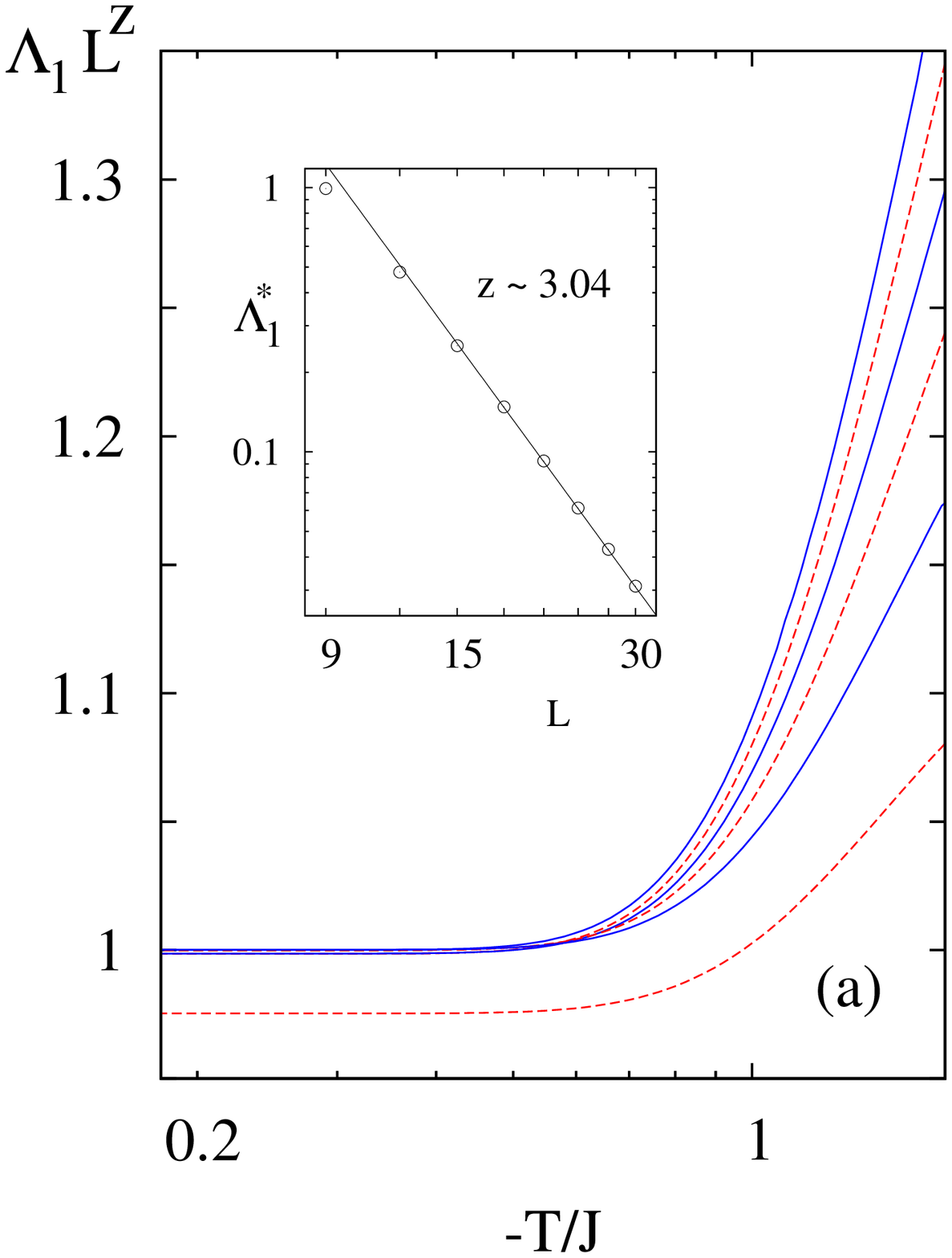}
\vskip -8.75cm
\hskip 7.4cm
\includegraphics[width=0.445\textwidth]{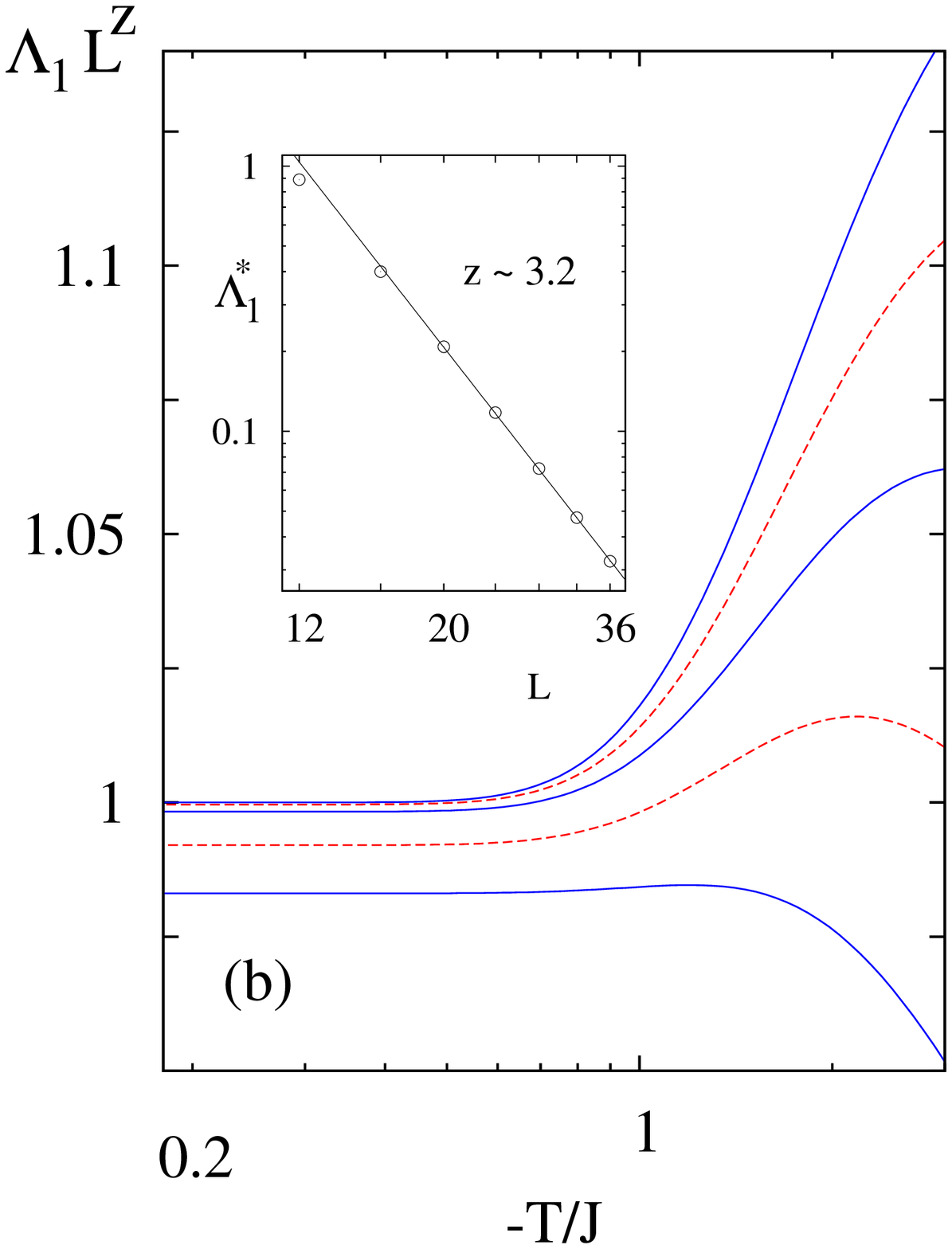}
\vskip 0.2cm
\caption{Scaling of normalized gaps [\,Eq.\,(\ref{normalization}), $L \propto k$\,] for NS 
spaces and $J < 0$. Solid and dashed lines  in upward order show alternately the cases of 
(a) $L = 15,\cdots,30$ for $k=3$, and (b) $L = 20,\cdots,36$ for $k=4$. The data collapse 
in the low-temperature regime was obtained using dynamic exponents read off from the 
slope of the insets (also see Table~\ref{tab1}). Just as in Fig.\,\ref{FSS}, for displaying 
convenience the vertical scales of main panels were respectively normalized by factors 
$30^z\,\Lambda_1^{\!^*}(30)$ and $36^z\,\Lambda_1^{\!^*}(36)$. The $\Lambda_1
\!$ plateaus are consistent with the Arrhenius regimes referred to in the text.}
\label{AF-FSS}
\end{figure}

At infinitesimal temperatures \cite{note2} and independently of the system size, the 
activation energies needed for these pair annihilations introduce divergences in the 
actual relaxation times via the Arrhenius factors $\propto e^{-4 J/T}$ mentioned in 
Sec.\,I. In estimating dynamic exponents from exact diagonalizations it is then natural to 
use the finite-size scaling hypothesis referred to in Eq.\,(\ref{normal}), which in terms of  
'normalized' spectral gaps reads
\begin{equation}
\label{normalization}
\Lambda_1^{\!^*} (L) \equiv \lim_{T \to 0^+}e^{\,-4 J/ T } \,\lambda_1 (L)\,
\propto L^{-z}.
\end{equation}
However, as temperature is lowered the spacing of the low-lying levels of the evolution 
operator gets arbitrarily small, so in practice it turns out that the pace of the Lanczos 
convergence becomes prohibitively slow for $T/ \vert J\vert\alt 0.15$. Nevertheless, a 
clear saturation trend already shows up for our lowest accessible $T$'s, thus signaling 
the entrance to the Arrhenius regime, and within which the normalized gaps are then 
scaled with the system size. This is exhibited in Figs.\,\ref{AF-FSS}(a) and \ref{AF-FSS}(b) 
for the NS sectors of $k=3$ and 4 respectively. Note that even a slight deviation from the 
conjectured energy barriers would result in strong departures from those $\Lambda_1\!$ 
plateaus. More importantly, the finite-size decay of these quantities within that region is 
indicative of dynamic exponents rather different from the diffusive ones obtained under F 
couplings. Their values are estimated by the slopes shown in the insets, and produce the 
collapse of larger size data in nearing the Arrhenius regime. A more detailed trend of size 
effects on these subdiffusive exponents is provided by the sequence of effective 
approximants
\begin{equation}
\label{approx}
Z_L (k) = \frac{\ln\big[\,\Lambda_1^{\!^*} (L) / \Lambda_1^{\!^*} (L-k)\,\big]}
{\ln\big[\,(L-k)/L\,\big]}\,,
\end{equation}
which simply derives a measure of $z (k)$ from the gaps of successive chain lengths 
$\propto k$. In Table~\ref{tab1} we list our higher approximants which happen to come 
out as forming sequences of upper and lower bounds for $z(3)$ and $z(4)$ respectively.
\begin{table}[htbp]
\begin{center}
\begin{tabular} {c  c  c}
\hline \hline
$L/k$ & \hskip 0.5cm  $Z_L\,(k=3)$  &  \hskip 0.4cm $Z_L\,(k=4)$
\vspace{0.1cm}
\\  
\hline 
\;7 &  \hskip 0.5cm 3.0713  & \hskip 0.3cm 3.1558
\\
\;8 & \hskip 0.5cm  3.0662 & \hskip 0.3cm 3.1887
\\
\;9 &  \hskip 0.5cm 3.0577  & \hskip 0.3cm 3.2117
\\
10 & \hskip 0.5cm  3.0431   & 
\vspace{0.1cm}
\\
\hline \hline
\end{tabular}
\end{center}
\caption{Convergence of dynamic exponents as resulting from the slopes between
larger available sizes in the insets of Fig.\,\ref{AF-FSS}.}
\label{tab1}
\end{table}
The case of trimers seems to converge towards a Lifschitz-Slyozov behavior [\,$\xi (t) 
\propto t^{1/3}$\,], similar to that encountered in the ferromagnetic 1D Kawasaki 
dynamics \cite{Cornell} as well as in $d \ge 2$ by surface  dynamical arguments 
\cite{Huse}. As already pointed out, the connection to the Kawasaki dynamics stems
from its mapping to the dimer case, though with opposite coupling signs \cite{note1}. 
It is then interesting to check that either under AF or F couplings the dynamic exponents
of the trimer case ($z \simeq 3$ and 2 respectively), also  follow those of the corresponding 
dimer (Kawasaki) kinetics. This common behavior of dimers and trimers contrasts with that 
of the case of $J= 0$ where it was found that their autocorrelation functions in NS sectors 
decay in time with different power laws \cite{Barma1,Barma2}. On the other hand, for 
$k=4$ the approximants of Table~\ref{tab1} suggest a slightly slower kinetics, possibly 
also implying non-universality on $k$ (contrariwise to the case $J > 0$ discussed in 
Sec.\,III\,A).

\subsection{Ground state characterization ($J < 0$)}

Turning to equilibrium at $T = 0^+$, as stressed by the end of Sec.\,II\,A the NS constraint 
impedes the ordering of AF configurations for $k \ge 3$ when $J < 0$. Instead, a highly 
degenerate structure emerges. It is similar to that of the metastable states described in 
Sec.\,III\,B except in that neither eventual adsorptions nor desorptions of $k$-mers would 
give rise to more than $2 k$ parallel spins. Hence, subsequent diffusion of kinks like those 
schematized in Fig.\,\ref{decay} could not meet the conditions to reduce the energy further
(see discussion below).

\begin{figure}[htbp]
\hskip -7.8cm
\includegraphics[width=0.42\textwidth]{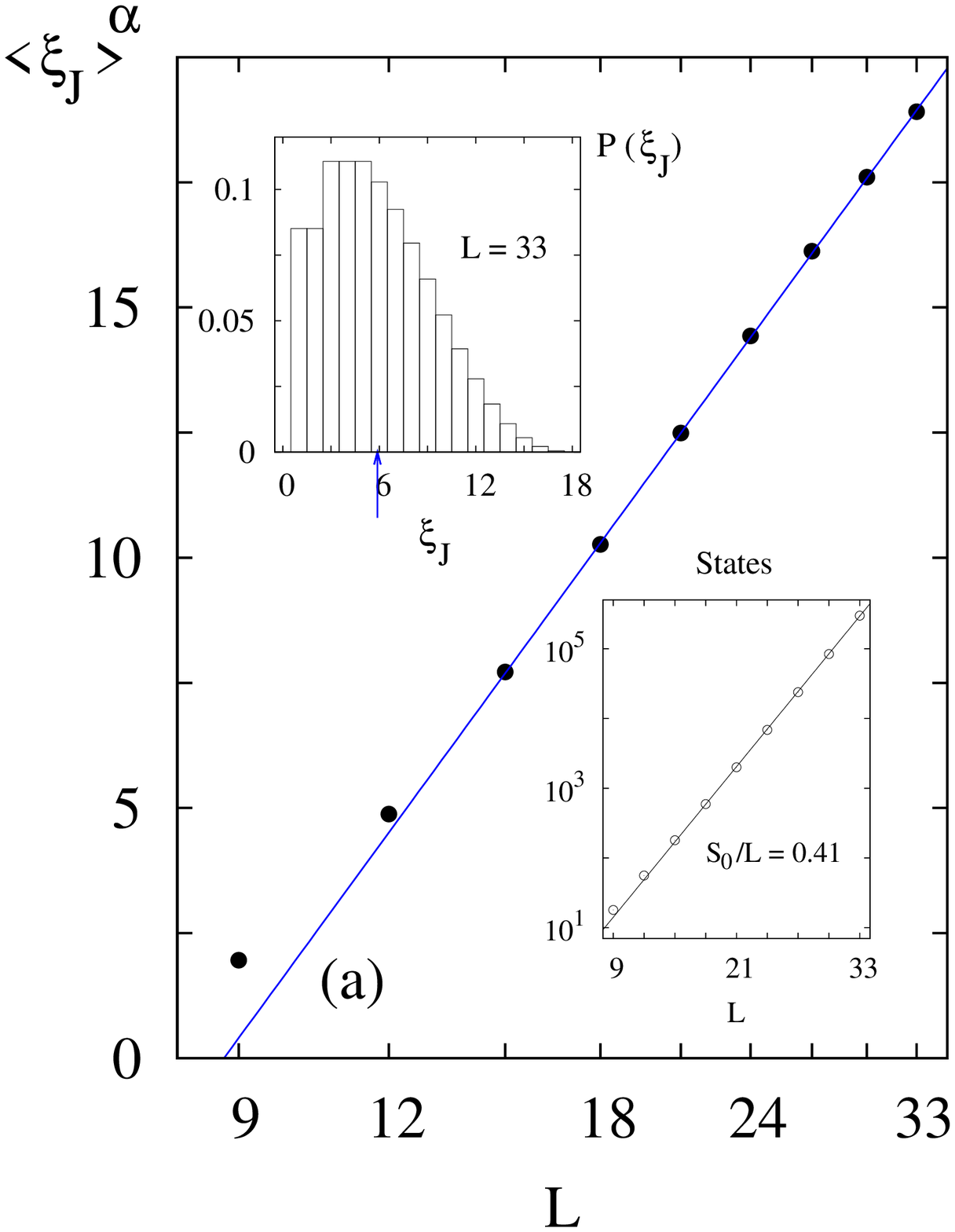}
\vskip -8.74cm
\hskip 7.4cm
\includegraphics[width=0.42\textwidth]{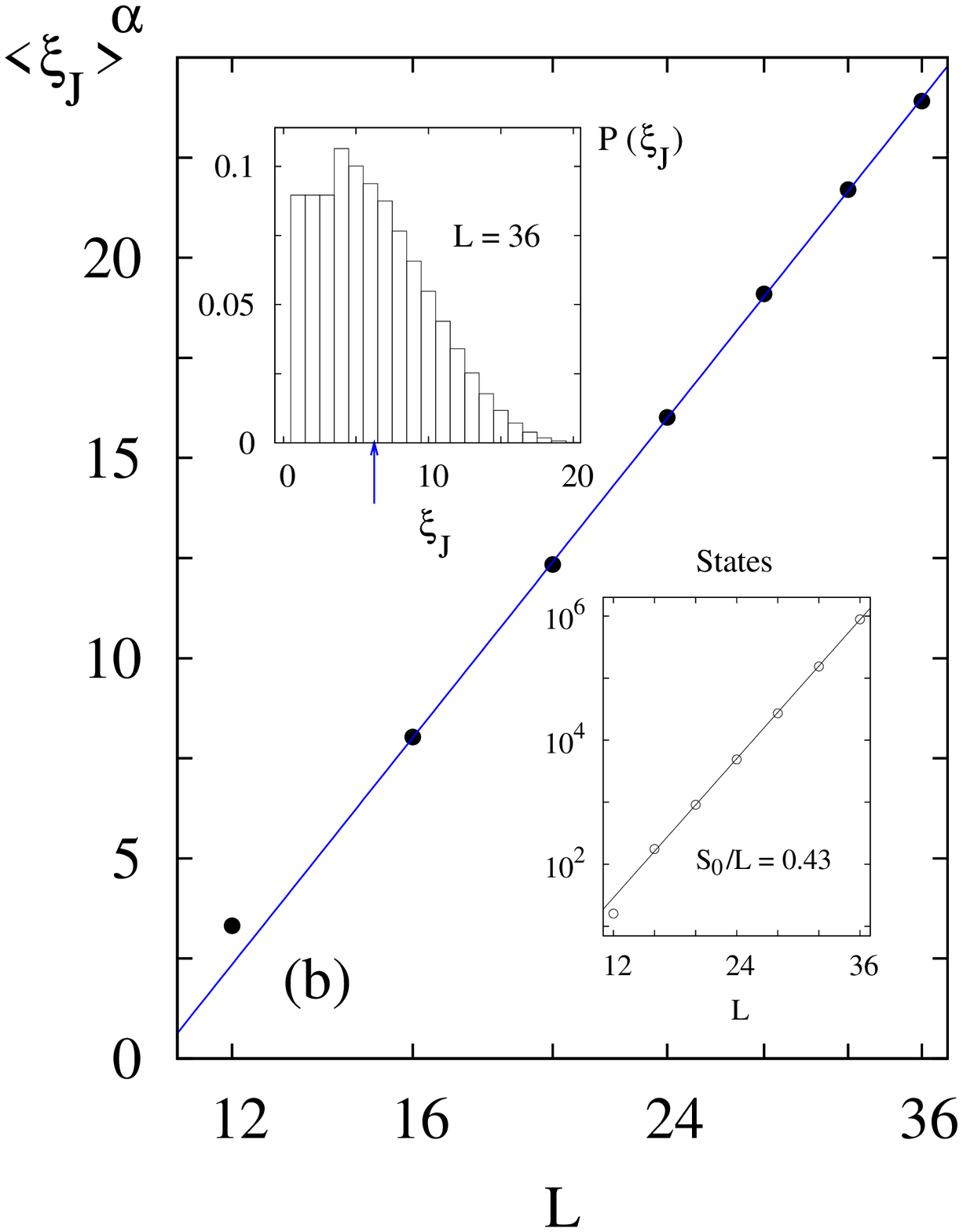}
\vskip 0.1cm
\caption{Main panels: logarithmic growth of average jamming lengths $\xi_{_J}$ with the 
system size ($L \propto k$) in the ground states of NS sectors with $J < 0$ for (a) $k=3$ 
($\alpha\simeq 1.6$), and (b) $k=4$ ($\alpha\simeq 1.7$). The distribution of these length 
scales in our largest accessible chains are shown in the upper histograms. Vertical arrows 
signal the mean values of $\xi_{_J}$. Lower insets illustrate the exponential degeneracy of 
the corresponding ground states. In turn, the  slopes indicate a residual entropy per site 
$S_0/L \sim 0.41$ for $k = 3$, and $\sim 0.43$ for $k = 4$.}
\label{jamming}
\end{figure}

When it comes to jamming scales, i.e. distances between flippable $k$-mers or lengths 
$\xi_{_J}$ through which there can be at most $k-1$ consecutive parallel spins, we have 
estimated their growth with $L$ by exact enumerations of ground states in the lattice sizes 
at reach.  Since at finite temperatures the dynamics is ergodic on each subspace, those 
regions can not remain jammed at all times. However, in the limit of $T \to  0^+$ note that 
the activation barriers referred to in Sec.\,III\,B allow those jammed regions to persist for 
arbitrarily large times in turn $\propto e^{4 \vert J \vert /T}$. The data exhibited in Fig.\,
\ref{jamming} are consistent with an average jamming length growing as $\langle \xi_{_J} 
\rangle \propto (\ln L)^{1/\alpha}$ with $\alpha \sim 1.6$ for $k=3$, and $\alpha \sim 1.7$ 
for $k=4$, thus suggesting a small $k$-mer density (active regions). Moreover, the
wide-tailed distributions of these lengths (upper insets), indicate the abundance of much 
broader jamming scales, a situation which is in part reminiscent of that found in RSA 
processes (cf. Sec.\,I). 
As for the high degeneracy of these states, the lower insets of Fig.\,\ref{jamming} clearly 
evidence an exponential growth of their number with the system size, which implies a
non-vanishing entropy in the low-temperature limit. In all examined cases the minimum 
energy is realized by configurations containing $2 \left(\frac{L}{k}-1\right)$ kinks subject 
to the NS constraint.

Owing to the Arrhenius barriers the sampling of ground states in larger chains would be 
hardly accessible to standard simulations. To bypass this problem, we just implement the 
decay pattern of the typical metastable configurations considered in Sec.\,III\,B, and in 
particular schematized for the case of trimers in Fig.\,5. Firstly, we flip a $k$-mer location {\it 
regardless} of its energy cost while checking that, as a result, at least $2k+1$ parallel spins 
are left (cf. Fig.\,\ref{decay}). Secondly, we allow the dynamics to proceed, but only through 
kink {\it diffusion} ($\Delta E = 0$), i.e. avoiding pair creation-annihilation events, until a 
group of $2k+3$  contiguous parallel spins shows up. As schematized in Fig.\,\ref{decay} 
this finally permits to reduce the original metastable energy by creating {\it two} kink pairs. 
This rearrangement process is then repeated until there is no longer any $k$-mer whose flip 
brings about at least $2k+1$ parallel spins. It is this latter feature what actually 
distinguishes a ground state configuration from a metastable one. Note that if the process 
were to continue, then subsequent kink diffusion would arrange intervals of at most $2k+2$ 
parallel spins, and so only {\it one} kink pair could be accommodated to compensate the 
energy excess $(-4 J)$ caused by the initial $k$-mer flip. As a result, the system would just 
be left in another ground state configuration. Thereafter the sampling continues but starting 
from other independent metastable state, in turn rapidly obtainable from a quench down to 
$T=0$. As a consistency check, it is worth mentioning that all ground state samples in the 
NS sector reached a maximum of $2 \left( \frac{L}{k}-1\right)$ kinks, which is also in line 
with that obtained by exact enumeration in much smaller sizes. 

\begin{figure}[htbp]
\hskip -7.8cm
\includegraphics[width=0.43\textwidth]{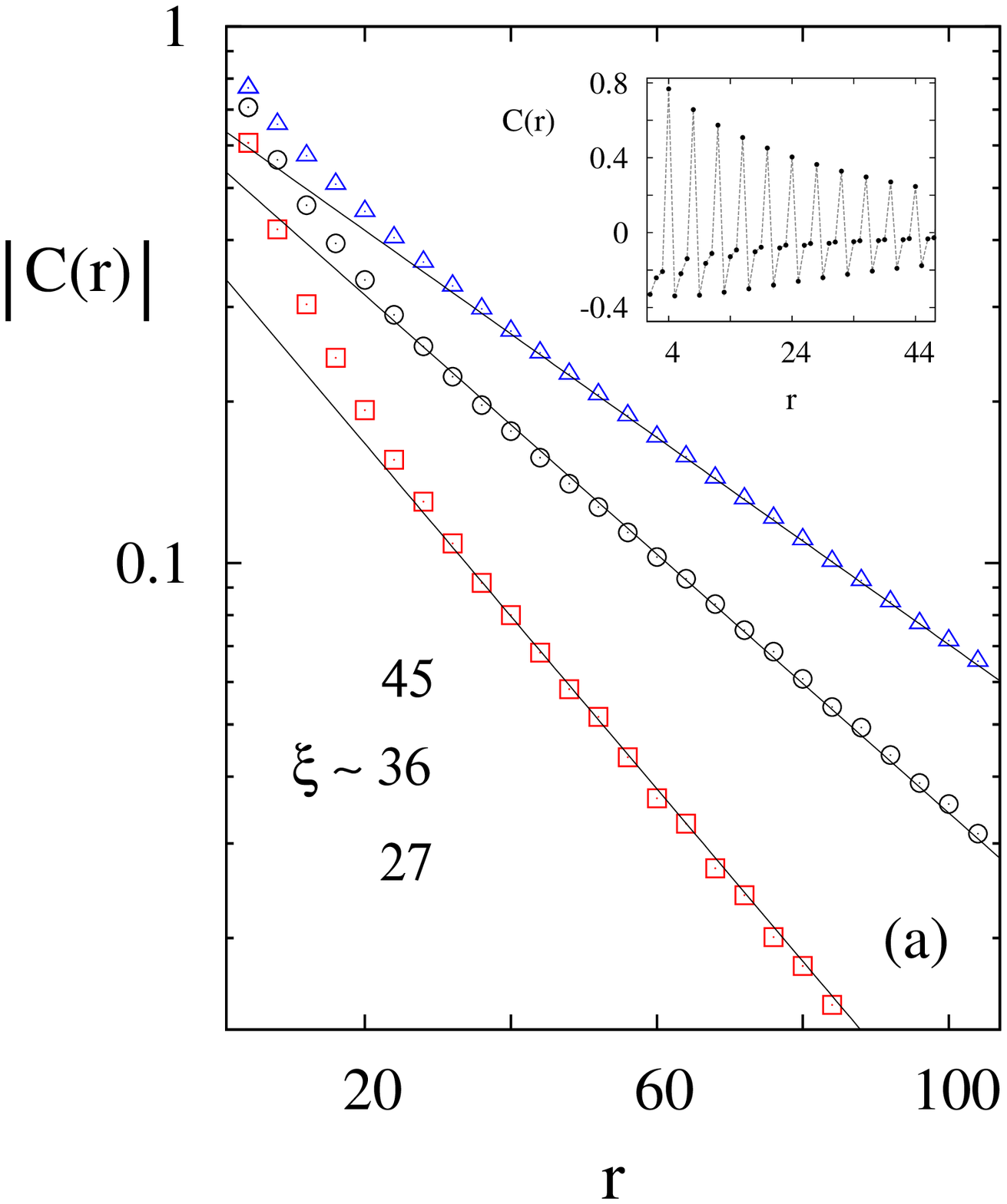}
\vskip -8.65cm
\hskip 7cm
\includegraphics[width=0.43\textwidth]{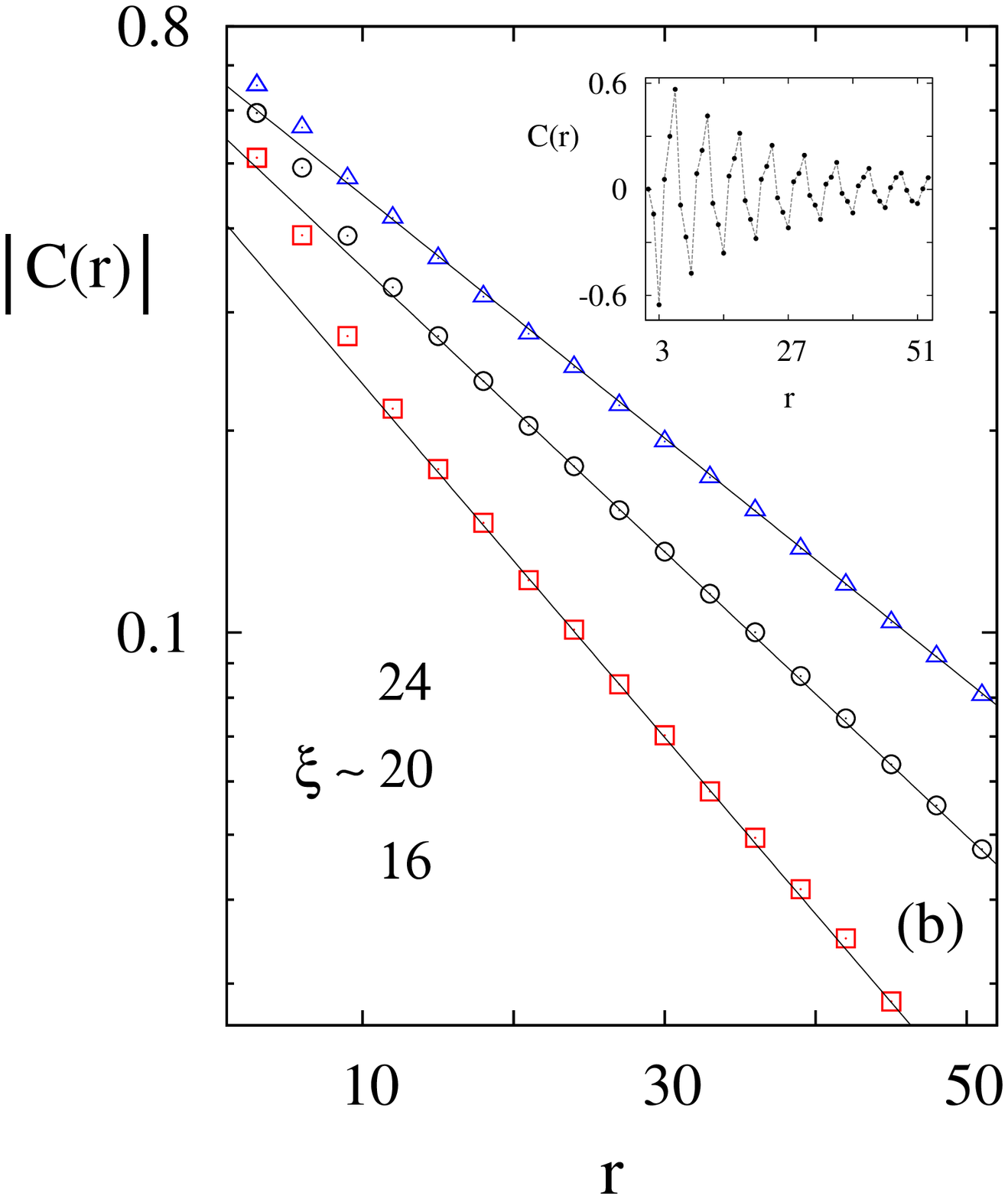}
\vskip 0.2cm
\caption{Spin-pair correlations in the ground states of NS sectors $(J < 0)$ obtained by the 
sampling procedure described in the text for (a) $k = 3$, and (b) $k = 4$. Squares, circles, 
and triangles stand respectively for $L = 600, 900, 1200$ in (a), and $L = 1800, 2400, 3000$ 
in (b). Error bars are smaller than twice the symbol sizes. The insets illustrate the cases of
largest $L$'s, whereas for displaying clarity only local extrema are shown in main panels. 
The inverse  slopes of solid lines are fitted with correlation lengths $\xi$ which grow with 
the chain size.}
\label{ground}
\end{figure}

The spin correlations $C (r) = \frac{1}{L} \sum_j \langle S_j\,S_{j+r} \rangle$ for $k=3$ 
and 4 resulting from this sampling scheme  are shown respectively in Figs.\,\ref{ground}(a) 
and \ref{ground}(b) after decaying from the metastable states generated by $\sim 10^4$ 
independent quenches. The rather large correlation lengths, in turn growing in proportion to 
the increase of the system size ($\Delta \xi\;\tilde{\propto}\;\Delta L$), presumably indicates 
long range order in the thermodynamic limit.
In that sense, it would be desirable to extend this picture to much larger chains, but there 
our sampling approach becomes progressively impractical. Finally, the insets exhibit the 
nontrivial forms which the NS constraint ends up imposing on these correlations, their 
oscillations for $k=3$ and 4 having respectively periods of four and six lattice units.

\section{Concluding discussion}

To summarize, we have studied the low-temperature and large-time dynamics of extended 
objects ($k$-mers) which reconstruct and interact while adsorbing/\,desorbing in
one-dimension. For $k \ge 3$ the notion of irreducible string \cite{Barma2} provided an
exhaustive description of the unusual manner in which the phase space is divided in sectors 
left invariant by the dynamics (Sec.\,II\,A). Although the number of such subspaces grows 
exponentially with the substrate size \cite{Barma1,Barma2}, we restricted the scope of this 
study to the so called null-string sector (containing the empty lattice configuration), both for 
computational ease and for being a common starting point in the context of cooperative AD 
processes \cite{Evans}. 

Thinking of these latter as multi-spin flip dynamics in Ising chains, we have constructed their 
evolution operators in the dual or kink representation of Sec.\,II\,B, by which we analyzed the 
scaling behavior of relaxation times in finite substrates. The resulting time scales were then 
read off from the spectral gaps obtained using standard recursive methods \cite{Lanczos}, 
ultimately enabling us to estimate dynamic exponents via the finite-size scaling hypothesis 
(\ref{normal}). In the case of F interactions (Sec.\,III\,A), the numerical matching of these 
gaps in the low-temperature limit with those of the rescaled Glauber operator ($L \to L/k$, 
Fig.\,\ref{four}a), might come as a bit of a surprise given the many-body correlations 
introduced by projectors (\ref{proj}) both in paring and hopping terms of Eqs.\,(\ref{pair}) 
and (\ref{diff}).  (Although in the limit of $T \to 0$ the creation of kink pairs is unlikely, the 
caging effect of those projectors on the remaining kinks is still important as not all of them 
necessarily move in the same sublattice. Hence, in the null sector, a strict analogy with the 
rescaled Glauber dynamics is not evident beyond two-kink excitations). The close overlap of 
those gaps near the critical regime thus strongly suggests a diffusive growth of relaxation 
times $\forall\, k \ge 1$ [\,Eq.\,(\ref{exact}), and data collapse of Fig.\,\ref{FSS}\,]. This 
was also corroborated by simulated quenches of spin-pair correlations (Fig.\,\ref{four}b), 
in all cases exhibiting ferromagnetic lengths which spread as $\sim k\,t^{1/2}$. Also upon 
normalizing all pair distances as in Fig.\,\ref{four}a ($r \to r/k$), these correlations were 
made to collapse into a single scaling function but different from that of the Glauber or 
monomer case \cite{Bray}. Whether this is due to the absence of conserved quantities, such 
as the irreducible strings or the sublattice magnetization differences of Eq.\,(\ref{m-diff}), 
remains an open issue. Let us add that this also might be an outcome of matrix elements 
of pair correlators being very different for $k > 1$ in the eigenstate basis of the evolution 
operator.

Owing to the metastable states appearing in the AF situation (Sec.\,III\,B), simulated 
quenches become impractical near the Arrhenius regime, so we contented ourselves with 
the above finite-size scaling methods, this time applied to the normalized gaps of Eq.\,(19). 
The scaling plots of these latter (Fig.\,\ref{AF-FSS}), as well as the sequence of our higher 
approximants [\,Eq.\,(\ref{approx}) and Table~\ref{tab1}\,], are indicative of subdiffusive 
dynamic exponents well apart from $z = 2$. For $k=3$ these seem to belong to the 
Lifschitz-Slyozov universality class \cite{Huse} characteristic of the ferromagnetic Kawasaki 
dynamics \cite{Cornell} (in this context, formally analogous to that of $k=2$ under $J < 0$), 
although for $k = 4$ the approximants of Table~\ref{tab1} suggest a convergence towards 
a slightly slower dynamics. This nonuniversal aspect clearly deserves further verifications 
in larger sizes, but already the next effective exponent ($Z_{40}$) would involve
diagonalizations in spaces of $\sim 5.02 \times 10^7$ dimensions.

With further regard to nonuniversality issues, it would be relevant to extend this study to 
the dynamics of string sectors with finite density of irreducible characters, such as those
considered in the noninteracting case of Ref.\,\cite{Barma2}. There it was found that
autocorrelation functions exhibit a wide diversity in the manner in which these decay in 
time, depending on the studied sector. In our case, preliminary diagonalizations for $J > 0$ 
in similar sectors however suggest that the diffusive picture found in Sec.\,III\,A still holds, 
though an understanding of the emergent low-temperature phases (also highly degenerate) 
would require further investigations.

In the null string subspace with $J < 0$ (where AF ordering is unattainable for $k >2$), that 
latter aspect was addressed in Sec.\,III\,C by exact enumerations in small chains alongside 
with the sampling of ground states in larger substrates. The former approach revealed both 
the appearance of growing jammed scales and finite residual entropies (Fig.\,\ref{jamming}), 
whereas the latter one \,-implemented by exploiting the decay pattern of Fig.\,\ref{decay}-\, 
disclosed nontrivial spin-pair correlations presumably long ranged in the thermodynamic 
limit (Fig.\,\ref{ground}). As an open problem also remains to elucidate to what extent these 
features, as well as the dynamic ones of Sec.\,III\,B, might be affected by the inclusion of a 
small magnetic field, i.e.  a monomer chemical potential slightly apart from that set by their 
couplings.

\section*{Acknowledgments}

We thank D. Dhar for helpful correspondence with observations and remarks. This work was 
partially supported by PIP 2012-0747 CONICET and PICT 2012-1724 ANPCyT.



\begin{thebibliography}{99}
\bibitem{note1} This just follows from the spin mapping $S_{2i} \to -S_{2i}$, say 
at even locations, while leaving the odd spins unchanged. Then  $S_i\,S_{i+1} \to
- S_i\,S_{i+1}$ whereas dimer deposition-evaporation is represented as a spin 
exchange between neighboring sites.
\bibitem{note2} At $T=0$ the dynamics rapidly gets stuck in metastable states, thus 
preventing the system to reach equilibrium.
\bibitem{note3} However, if the targeted $k$-mer happens to be contiguous to another 
one, then at least a third $k$-mer location is needed.
\end{thebibliography}
\end{document}